\documentclass[aps,prl,amsmath,amssymb,twocolumn,groupedaddress]{revtex4-1}

\usepackage{graphicx}  
\usepackage{color}
\usepackage{amssymb}   
\usepackage{amsmath}
\usepackage{epstopdf}
\usepackage{natbib}
\usepackage{hyperref}  
\usepackage{bm}		   
\usepackage{braket}
\usepackage{gensymb}
\usepackage{bbold}

\begin{document}
\title{Giant nonlinear Hall effect in strained twisted bilayer graphene}

\author{Cheng-Ping Zhang$^{1}$}
\author{Jiewen Xiao$^{2}$}
\author{Benjamin T. Zhou$^{1}$}
\author{Jin-Xin Hu$^{1}$}
\author{Ying-Ming Xie$^{1}$}
\author{Binghai Yan$^{2}$} \thanks{Correspondence author: binghai.yan@weizmann.ac.il}
\author{K. T. Law$^{1}$} \thanks{Correspondence author: phlaw@ust.hk}

\affiliation{$^{1}$Department of Physics, Hong Kong University of Science and Technology, Clear Water Bay, Hong Kong, China}
\affiliation{$^{2}$Department of Condensed Matter Physics, Weizmann Institute of Science, Rehovot, 7610001, Israel}

\begin{abstract}
Recent studies have shown that moir\'{e} flat bands in a twisted bilayer graphene(TBG) can acquire nontrivial Berry curvatures when aligned with hexagonal boron nitride substrate \cite{he2020giant, liu2020anomalous}, which can be manifested as a correlated Chern insulator near the 3/4 filling \cite{sharpe2019emergent, serlin2020intrinsic}. In this work, we show that the large Berry curvatures in the moir\'{e} bands lead to strong nonlinear Hall(NLH) effect in a strained TBG with general filling factors. Under a weak uniaxial strain $\sim 0.1\%$, the Berry curvature dipole which characterizes the nonlinear Hall response can be as large as $\sim 200 \; \text{\AA}$, exceeding the values of all previously known nonlinear Hall materials \cite{ma2019observation, kang2019nonlinear, you2018berry, zhang2018electrically, du2018band, battilomo2019berry, son2019strain, zhou2020highly, hu2020nonlinear, huang2020giant} by two orders of magnitude. The dependence of the giant NLH effect as a function of electric gating, strain and twist angle is further investigated systematically. Importantly, we point out that the giant NLH effect appears generically for twist angle near the magic angle due to the strong susceptibility of nearly flat moir\'{e} bands to symmetry breaking induced by strains. Our results establish TBG as a practical platform for tunable NLH effect and novel transport phenomena driven by nontrivial Berry phases. 
\end{abstract}

\maketitle

\begin{figure}
\centering
\includegraphics[width=3.5in]{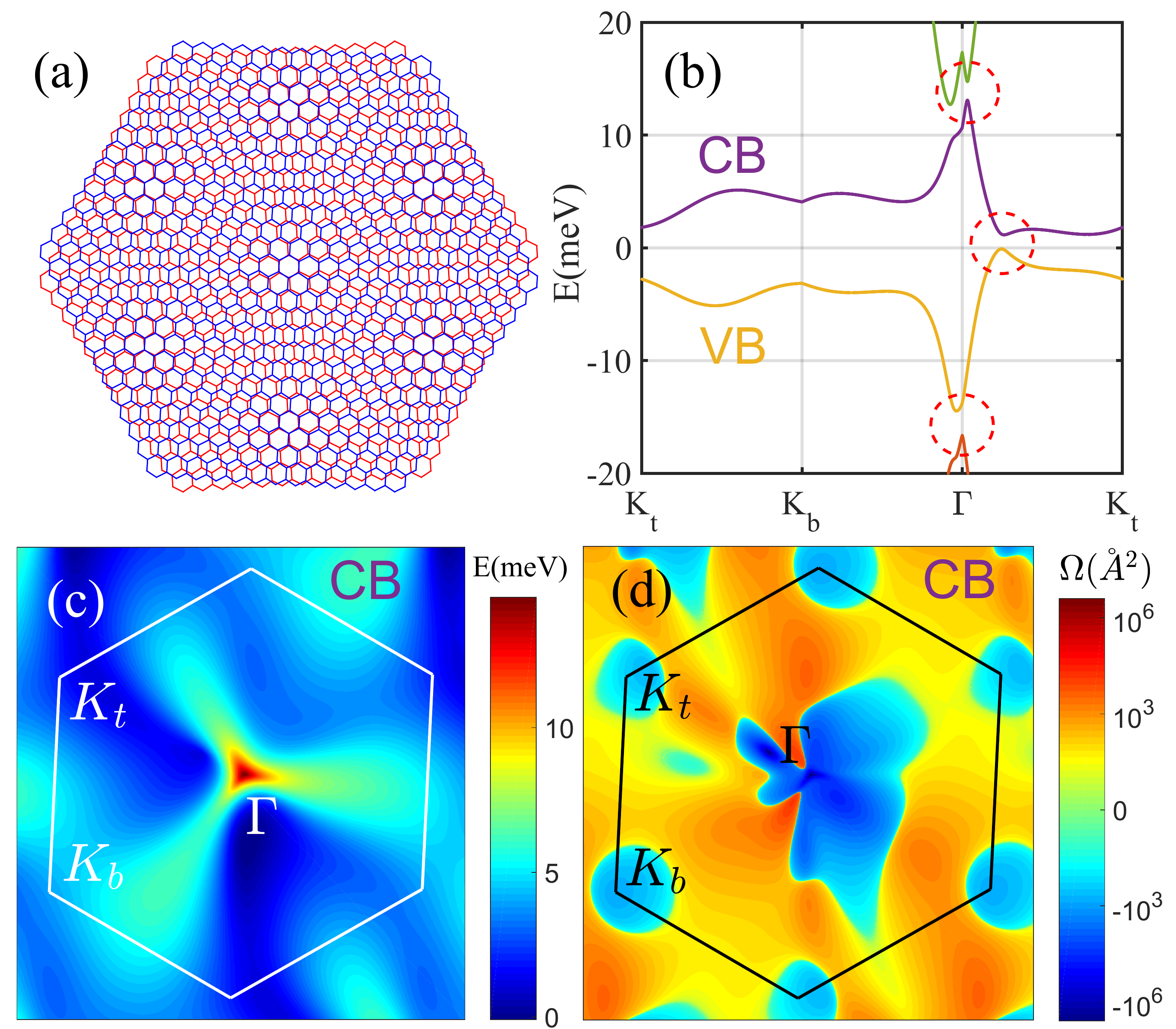}
\caption{(a) A schematic figure of TBG moir\'e superlattice at
a twist angle of 7$\degree$. (b) moir\'e bands near valley $\xi = +1$ of TBG at a twist angle $\theta = 1.2 \degree $, with strain $\mathcal{E} = 0.1 \%$ applied along the zigzag direction of the bottom layer. Zero energy is chosen in the middle of the band gap when strain is absent. The conduction band(CB) is in purple and the valence band(VB) is in yellow. Band anti-crossings happen near the band edges, indicated by red dashed circles, where large Berry curvature is found. (c)The energy dispersion of the conduction band in the whole moir\'e Brillouin zone. (d)Berry curvature of the conduction band, shown in logarithmic scale. Large negative Berry curvature in the order $\sim 10^6 \; {\text{\AA}^2}$ is found near the band anti-crossings.}
\label{FIG01}
\end{figure}

\emph{Introduction.}---Recently, the study of exotic properties of magic-angle twisted bilayer graphene(TBG) has become a major topic in condensed matter physics \cite{dos2007graphene, bistritzer2011moire, cao2018correlated, cao2018unconventional, yankowitz2019tuning, koshino2018maximally, kang2018symmetry, po2018origin, tarnopolsky2019origin, liu2019pseudo, song2019all}. A TBG is formed by two atomic sheets of graphene with a relative small twist angle $\theta$, as shown in Fig.\ref{FIG01}(a). The stacking of the two slightly misoriented graphene sheets leads to the formation of a long-period moir\'{e} pattern, which significantly reduces the Fermi velocity at the Dirac points. In particular, as the twist angle decreases, two nearly flat moir\'{e} bands emerge at a series of magic angles \cite{bistritzer2011moire} where electron-electron interactions become important. The recent observation of superconductivity and possible correlated insulating phase near half filling in magic-angle TBGs \cite{cao2018correlated, cao2018unconventional}, which are believed to be driven by strong electron correlations, has inspired intensive recent theoretical and experimental studies to explore exotic correlated phases in twisted graphene systems \cite{xu2018topological, xu2018kekule, jiang2019charge, kerelsky2019maximized, xie2019spectroscopic, choi2019electronic, cao2020nematicity, wu2018theory, xie2020nature, isobe2018unconventional, lian2019twisted, gonzalez2019kohn, liu2019nematic}.

Besides the progress in the study of correlated physics, recent experiments have also observed ferromagnetism and quantum anomalous Hall effect in magic-angle TBGs with 3/4 filling \cite{sharpe2019emergent, serlin2020intrinsic}, suggesting nontrivial topological properties of the moir\'{e} bands. The observation supports the theoretical proposal that strong electron-electron interactions spontaneously lift the spin and valley degeneracies and result in fully filled moir\'{e} bands with nonzero Chern numbers \cite{zhang2019nearly, zhang2019twisted}. Surprisingly, the direction of magnetization and Hall currents were found to be easily switchable upon applying a small DC current of $nA$ scale through the TBG samples. Such current-induced magnetic switching was recently explained to arise from the giant orbital magneto-electric effect in TBG \cite{he2020giant}, in which nontrivial Berry curvatures \cite{he2020giant, liu2020anomalous, liu2019nematic, xie2020nature} together with three-fold ($C_3$) symmetry breaking introduced by the hexagonal boron nitride ($h$-BN) substrate allow currents in TBGs to carry giant orbital magnetic moments, which enables the switching of the total magnetization.  

In analogy to the anomalous Hall effect in magnetic materials, the current-induced orbital magnetization in TBGs can combine again with the applied current to induce Hall currents via the mechanism known as the nonlinear Hall(NLH) effect \cite{sodemann2015quantum}. In particular, as the AB sublattice and three-fold symmetry breaking from the $h$-BN substrate reduces the point group of TBG from $D_6$ to $C_1$, a nonzero Berry curvature dipole which characterizes the NLH response is allowed. While the recent observation of NLH effects in WTe$_2$ \cite{ma2019observation, kang2019nonlinear} has motivated on-going efforts to explore novel platforms \cite{facio2018strongly, battilomo2019berry, son2019strain, zhou2020highly} for NLH physics as well as its potential applications, the study of NLH effects in twisted materials has just started lately \cite{hu2020nonlinear, huang2020giant}.

In this work, we show that the large Berry curvatures in the moir\'{e} bands of a strained TBG aligned with $h$-BN substrate give rise to a strong NLH effect for general filling factors. In particular, under a weak uniaxial strain $\sim 0.1\%$, the optimal Berry curvature dipole can be as large as $\sim 200 \; \text{\AA}$ and exceed the values of all previously known nonlinear Hall materials by two orders of magnitude \cite{ma2019observation, kang2019nonlinear, you2018berry, zhang2018electrically, du2018band, battilomo2019berry, son2019strain, zhou2020highly, hu2020nonlinear, huang2020giant}. Importantly, the giant NLH happens generically for twist angles close to the first magic angle $\sim 1.1 ^{\circ}$ where the flat moir\'{e} bands enables a dramatic symmetry breaking from strains and creates a large Berry dipole. Our results establish TBG as a practical platform for tunable NLH effect and other novel transport phenomena driven by Berry dipole physics \cite{zeng2019nonlinear, zeng2019wiedemann, yu2019topological}. 

\emph{Continuum model of strained TBG.}---At a small twist angle $\theta$, the low-energy physics of a TBG can be described by a continuum model formed by massless Dirac fermions in each layer \cite{dos2007graphene, bistritzer2011moire}. In the layer basis, the effective Hamiltonian for valley $\xi = \pm$ can be written as:
\begin{equation}
\hat{H}_{\xi}=\begin{pmatrix}
\hat{H}_{b,\xi} & T_{\xi}(\bm{r})\\
T_{\xi}^{\dagger}(\bm{r}) & \hat{H}_{t,\xi}
\end{pmatrix}
\label{eq:continumm},
\end{equation}
where $t$ ($b$) labels the top (bottom) layer, which is rotated by $+(-) \frac{\theta}{2}$ about the $z$-axis. The Hamiltonian of each monolayer describes the massless Dirac fermion 
\begin{equation}
H_{t/b, \xi} = \hbar v_{F} \hat R _{\mp \frac{\theta}{2}} \bm{q} \cdot (\xi \sigma_x, \sigma_y),
\end{equation}
where $ \hbar v_{F} = 5.96 \; \text{eV} \cdot \text{\AA}$ is the original Fermi velocity \cite{neto2009electronic}, and $\hat R_{\theta}$ is the rotation operator. The momentum $\bm{q} = \bm{k} - \bm{K}_{\xi}$ is defined relative to the original Brillouin zone corner $\bm{K}_{\xi}$, and $\bm{\sigma}$ are the Pauli matrices acting on the AB sublattice space. The interlayer hopping is
\begin{eqnarray}\nonumber
T_{\xi}(\bm{r}) &=&
\begin{pmatrix} 
u & u' \\
u' & u 
\end{pmatrix}
e^{-i \xi \bm{q_{b}}\cdot\bm{r}} +
\begin{pmatrix}
u & u' e^{-i\xi\phi} \\
u' e^{i\xi\phi} & u 
\end{pmatrix}
e^{-i \xi \bm{q_{tr}}\cdot\bm{r}} \\
&& +
\begin{pmatrix} 
u & u' e^{i\xi\phi} \\
u' e^{-i\xi\phi} & u
\end{pmatrix}
 e^{-i \xi \bm{q_{tl}}\cdot\bm{r}},
\end{eqnarray}
with $\phi = 2 \pi / 3$. For simplicity, we use tunneling parameters $u = u' = 110$ meV \cite{bistritzer2011moire} in our calculations. The effect of lattice corrugation in
TBG can be included by choosing different tunneling amplitudes $u$ and $u'$ \cite{koshino2018maximally}, which do not affect our results in a qualitative way (see Supplementary Material \cite{NoteX} for details). $\bm{q}_{b}$, $\bm{q}_{tr}$,  $\bm{q}_{tl}$ denote the momentum transfers of the three hopping processes \cite{bistritzer2011moire, NoteX}.

A pristine TBG respects the $D_6$ symmetry, which consists of a $C_{6}$ symmetry of each graphene layer and a $C_{2x}$ symmetry swapping the two layers, as shown in Fig.\ref{FIG01}(a). Dirac crossings at the moir\'{e} Brillouin zone corners are protected by the composite $C_{2z}\mathcal{T}$ symmetry \cite{po2018origin}. Furthermore, the $C_{2z}$ symmetry requires the Berry curvature of a band indexed by $n$ to be even in momentum space: $\Omega_{n,-\xi}(-\bm{q}) = \Omega_{n,\xi}(\bm{q})$, while time-reversal ($\mathcal{T}$) symmetry requires $\Omega_{n}$ to be odd: $\Omega_{n,-\xi}(-\bm{q}) = - \Omega_{n,\xi}(\bm{q})$. Thus, the composite $C_{2z}\mathcal{T}$ symmetry forces the Berry curvature to vanish throughout the entire moir\'{e} Brillouin zone \cite{wu2018theory, xie2020nature}.

However, under realistic experimental conditions, aligning a TBG with $h$-BN substrate induces a staggered potential on the bottom graphene layer \cite{zhang2019twisted, he2020giant, liu2020anomalous}, which breaks the $C_{2x}$ and $C_{2z}$ symmetries and reduces the symmetry to $C_3$. Particularly, upon breaking the $C_{2z}$ (or equivalently, the AB sublattice symmetry), the Dirac crossings at the zone corners are gapped out, and non-trivial Berry curvatures emerge. Moreover, as strains are found to be prevalent in twisted materials \cite{kerelsky2019maximized,choi2019electronic, xie2019spectroscopic, huang2020giant}, the remaining $C_3$ symmetry is expected to be broken in realistic TBG samples as well, which allows a nonzero Berry curvature dipole and NLH effect is expected to be manifested experimentally.

To include strain effects in TBG, we introduce the linear strain tensor $\bm{\mathcal{E}}$, which transforms a general coordinate ($\bm{r},\bm{k}$) in phase space as \cite{bi2019designing}:
\begin{eqnarray}
\bm{r'} &=& (\mathbb{1} + \bm{\mathcal{E}}) \bm{r}, \\\nonumber
\bm{k'} &=& (\mathbb{1} + \bm{\mathcal{E}}^T)^{-1} \bm{k} \approx (\mathbb{1} - \bm{\mathcal{E}}^T) \bm{k}.
\end{eqnarray}
Without loss of generality, consider a uniaxial strain of magnitude $\mathcal{E}$ along the direction with angle $\varphi$ relative to the zigzag direction, the explicit form of the strain tensor is written as \cite{pereira2009tight, bi2019designing}:
\begin{eqnarray}\label{eq:strain}
\bm{\mathcal{E}} &=& \hat R _{\varphi}
\begin{pmatrix}
\mathcal{E} & 0 \\
0 & - \nu \mathcal{E}
\end{pmatrix}
\hat R^{-1}_{\varphi}\\\nonumber
&=& \mathcal{E}
\begin{pmatrix}
\cos ^2 \varphi - \nu \sin ^2 \varphi & (1 + \nu) \cos \varphi \sin \varphi \\
(1 + \nu) \cos \varphi \sin \varphi & - \nu \cos ^2 \varphi + \sin ^2 \varphi
\end{pmatrix},
\end{eqnarray}
with $\nu = 0.165$ \cite{pereira2009tight} being the Poisson ratio of graphene.

For the bottom layer Hamiltonian, the strain introduces an effective gauge field \cite{guinea2010energy}:
\begin{equation}
\bm{A} = \frac{\beta}{d}(\mathcal{E}_{xx}-\mathcal{E}_{yy}, -2\mathcal{E}_{xy}),
\end{equation}
with coefficient $\beta=1.57$ \cite{guinea2010energy}, and $d = 1.42 \; \text{\AA}$ \cite{neto2009electronic} is the carbon-carbon bond length, which shifts the positions of the bottom layer Dirac points to $ \bm{D}_{\xi} = (\mathbb{1} - \bm{\mathcal{E}}^T) \bm{K}_{b,\xi} - \xi \bm{A} $. Together with the staggered potential $\Delta \sigma_z$ induced by the $h$-BN substrate, the bottom layer Hamiltonian can be written as \cite{he2020giant}
\begin{equation}\label{eq:H_b}
H'_{b, \xi} = \hbar v_{F} \hat R _{ \frac{\theta}{2}} [(\mathbb{1} + \bm{\mathcal{E}}^T)\bm{q'} + \xi \bm{A}] \cdot (\xi \sigma_x, \sigma_y) + \Delta \sigma_z,
\end{equation}
with $\Delta = 17$ meV \cite{kim2018accurate}. Strain effects on interlayer terms are included in the change in momentum transfers in the interlayer hopping process (see Supplementary Material \cite{NoteX}).

The energy spectrum of TBG at twist angle $ \theta = 1.2 \degree $ under strain $\mathcal{E} = 0.1 \%$ along the zigzag direction is plotted in Fig.\ref{FIG01}(b). Note that the nonzero staggered potential $\Delta$ gaps out band crossings generically, \textit{i.e.}, a band anti-crossing occurs, with hybridization gaps on the order of several meVs (indicated by the red dashed circles in Fig.\ref{FIG01}(b)), where large Berry curvatures can emerge.

The energy dispersion of the conduction band in the entire moir\'e Brillouin zone is further depicted in Fig.\ref{FIG01}(c), which clearly demonstrates the breaking of $C_3$ symmetry under strains. The conduction band maximum is located near the $\Gamma$ point, and the band minimum is located near the $\Gamma - K_{t}$ lines, being consistent with Fig.\ref{FIG01}(b). To explicitly demonstrate the emergence of Berry curvatures near band anti-crossing points, we calculate the Berry curvature of the conduction band:
\begin{eqnarray}\label{eq:BC}
\bm{\Omega}_n = i \bra{\partial_{\bm{k}} u_n} \times \ket{\partial_{\bm{k}} u_{n}},
\end{eqnarray}
with $\ket{u_n}$ being the periodic part of the Bloch wave function at $\bm{k}$. The momentum-space profile of $\bm{\Omega}_n$ is shown in logarithmic scale in Fig.\ref{FIG01}(d), which clearly shows that large Berry curvatures of order $\sim 10^6 \; {\text{\AA}^2}$ appear at band anti-crossing points near $\Gamma$ and the $\Gamma - K_{t}$ lines. Notably, the breaking of $C_3$ due to strain results in a highly non-uniform Berry curvature profile in the moir\'{e} Brillouin zone (Fig.\ref{FIG01}(d)), which indicates the presence of a large Berry dipole and a strong NLH effect. 

\emph{Strain-gate map of giant NLH effect in near-magic-angle TBG.}---We first study the NLH effect in a near-magic-angle TBG with twist angle $\theta = 1.2 \degree $. By systematically investigating the strain and gating dependence of the Berry curvature dipole, we show that giant NLH effects characterized by Berry dipoles of order $\sim 200 \; \text{\AA}$ appear generically when the Fermi level is close to the band anti-crossing points in TBGs under a weak uniaxial strain $\sim 0.1 \%$.

In the NLH effect, application of an AC electric field $\bm{E}(t)=Re\{ \bm{\varepsilon} e^{i \omega t} \}$ of frequency $\omega$ induces a transverse Hall current $\bm{j}(t) = Re \{ \bm{j}^0 + \bm{j}^{2\omega} e^{2 i \omega t} \}$ with a rectified component $j^0_a = \chi_{abc} \varepsilon_b \varepsilon_c^*$ and a second-harmonic component $j^{2 \omega}_a = \chi_{abc} \varepsilon_b \varepsilon_c$ with frequency $2\omega$. Here, the nonlinear Hall susceptibility $\chi_{abc} = \frac{e^3 \tau}{2(1 + i \omega \tau)} \varepsilon_{adc} D_{bd}$ is characterized by the Berry curvature dipole $D_{bd}$ \cite{sodemann2015quantum}, $e$ is the electron charge, $\tau$ is the scattering time, $\varepsilon_{adc}$ is the Levi-Civita tensor. The Berry curvature dipole in 2D is given by \cite{sodemann2015quantum}:
\begin{equation}\label{eq:BCDipole}
D_{bd} = \sum_{n,\xi} \int_{\bm{k}} f_0 \partial _b \Omega_d .
\end{equation}
Here, the summation is over band index $n$ and valley index $\xi$, $\int_{\bm{k}}=\int d^2 k/(2\pi)^2$, and $\partial _b=\partial/\partial_{k_b}$. $ f_0$ is the equilibrium distribution function and $\Omega_d$ is the Berry curvature with $d=z$ in 2D. In the following, we use the notation $D_{b} \equiv D_{bz}$ ($b = x,y$) for the $x,y$-components of Berry dipole in 2D, where the $x(y)$-axis is defined as the angular bisector between the two zigzag(armchair) directions of top and bottom graphene layers. 

\begin{figure}
\centering
\includegraphics[width=3.5in]{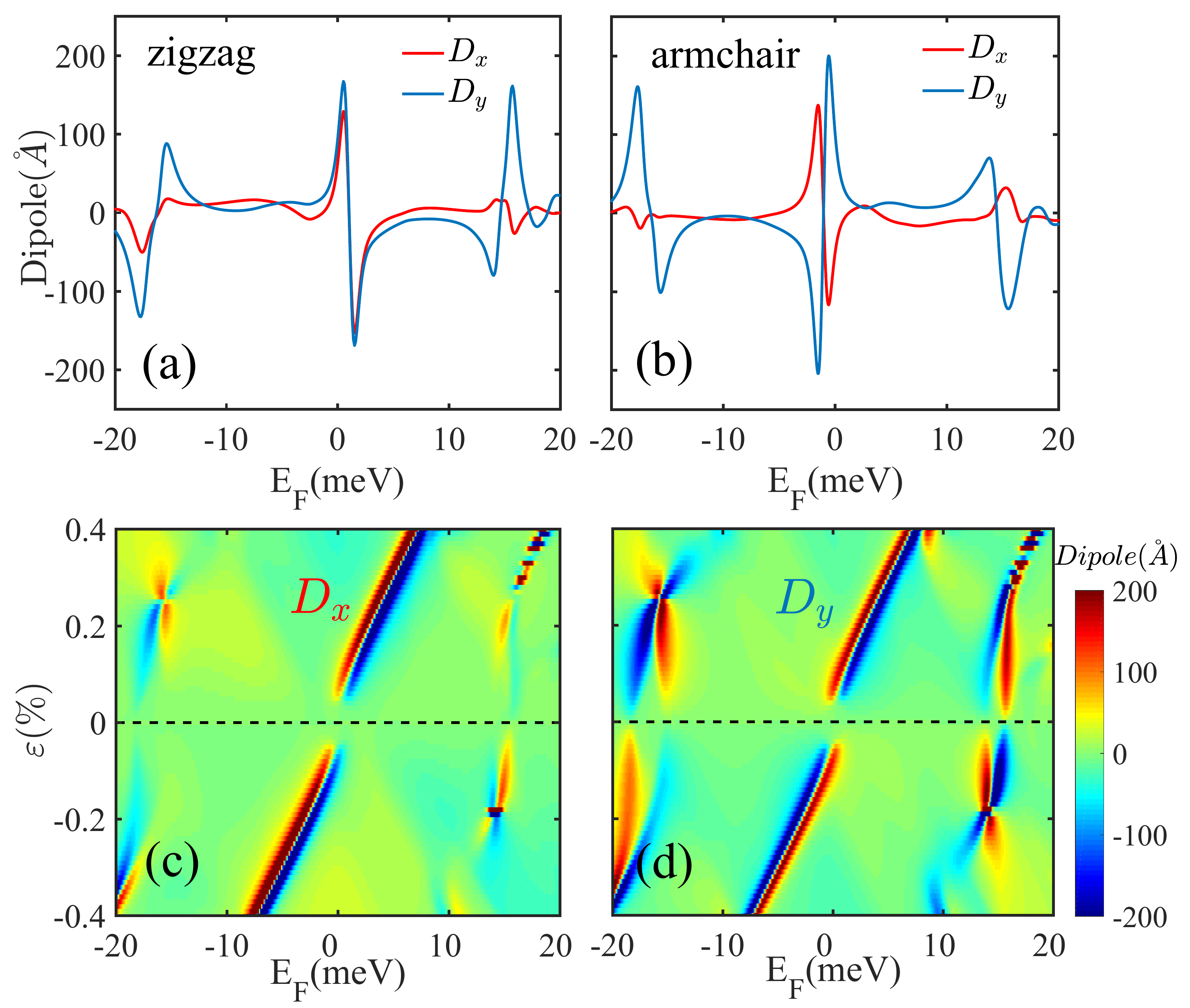}
\caption{(a) Berry curvature dipole components $D_x$ and $D_y$ at fixed strain $0.1\%$, which is applied along the zigzag direction($\varphi = 0 \degree$). The twist angle $\theta$ is fixed at 1.2$\degree$, and the temperature is set at 2 K. (b) Same parameters as in panel (a), with the strain applied along the armchair direction($\varphi = 90 \degree$). (c)-(d) Evolution of the Berry curvature dipole components $D_x$ and $D_y$, with respect to both the Fermi energy $E_F$ and the strain $\mathcal{E}$. Positive(negative) $\mathcal{E}$ represents tensile(compressive) strain. The parameters are the same as in panel (a).}
\label{FIG02}
\end{figure}

The gate dependence of $D_x, D_y$ at $\theta = 1.2 \degree$ under a fixed strain $\mathcal{E} = 0.1\%$ along the zigzag direction is shown in Fig.\ref{FIG02}(a), with a total factor of 4 accounting for spin and valley degeneracies. Evidently, optimal Berry curvature dipoles of order $\sim 200 \; \text{\AA}$ appear generically near the band anti-crossing points in Fig.\ref{FIG01}(b), which is two orders of magnitude larger than the optimal values $\sim 1 \; \text{\AA}$ observed in previous experiments \cite{ma2019observation, kang2019nonlinear, huang2020giant}.

A closer inspection into the gate dependence of $D_x, D_y$ reveals that the Berry dipoles generally changes sign when the anti-crossing points are accessed (indicated by red dashed circles in Fig.\ref{FIG01}(b)). This is due to the fact that the two bands near band anti-crossing points carry opposite Berry curvatures in general, and the Berry curvature dipole changes sign as $E_F$ goes across the band anti-crossing points. This gives rise to the dip-to-peak or peak-to-dip features in $D_x, D_y$ for $E_F$ in the ranges $-20 \sim -13$ meV, $-2 \sim 4$ meV, and $11 \sim 18$ meV, as shown in Fig.\ref{FIG02}(a). Similar features also appear for strain applied along the armchair direction (Fig.\ref{FIG02}(b)).

To demonstrate the crucial role of $C_3$ breaking in creating the nonzero Berry dipoles, we further calculate a complete $2D$ map of $D_x$ and $D_y$ as a function of both the Fermi energy $E_F$ and the uniaxial strain $\mathcal{E}$, as shown in Fig.\ref{FIG02}(c) and Fig.\ref{FIG02}(d) respectively. Clearly, in the absence of strain $\mathcal{E}= 0$, the $C_3$ symmetry forces the dipole to vanish in all ranges of $E_F$. As the strain is turned on, nonzero $D_x, D_y$ gradually emerges. Importantly, an experimentally relevant strain $\sim 0.1\%$ is already sufficient to induce a giant Berry curvature dipole, and we expect the NLH in TBG to be easily seen in nonlinear Hall experiments.

Notably, in the strain-gate maps of $D_x, D_y$ (Fig.\ref{FIG02}(c)-(d)), butterfly-like patterns generally appear near the conduction band maximum and valence band minimum with a critical strain $\mathcal{E}_c \sim \pm 0.2 \%$. This feature originates from a strain-induced topological band inversion which interchanges the Berry curvature signs of neighboring bands and involves a change in the valley Chern number of the conduction(valence) band from $-1 (+1)$ to $0$ (see Supplementary Material \cite{NoteX} for details). However, before the band inversion happens at $\mathcal{E}_c$, a strain with $\mathcal{E} < \mathcal{E}_c$ can cause the conduction and valence band to overlap and the TBG becomes semimetallic at the charge neutrality point, and the strain-induced topological band inversion is not likely to be manifested in the anomalous Hall conductance due to the coexistence of bulk channels. This is consistent with previous observations that the quantization of the anomalous Hall conductance near 3/4 filling is sensitive to $C_3$ breaking from strains \cite{serlin2020intrinsic, he2020giant, liu2019nematic}. Therefore, the butterfly patterns in Fig.\ref{FIG02}(c)-(d) can serve as a distinctive signature of the nontrivial topological band inversion induced by strain. 

\emph{ Role of magic-angle band flatness in giant NLH effect.}---Flatness of the moir\'{e} bands near magic angles in TBG has been shown to be important for exotic correlated physics at special filling factors such as 1/2 \cite{cao2018correlated, cao2018unconventional} and 3/4 \cite{yankowitz2019tuning,sharpe2019emergent,serlin2020intrinsic}. It has also been shown that the large density of states due to band flatness is responsible for the giant orbital magneto-electric effect in TBG \cite{he2020giant}. Here, we point out that the flatness of moir\'{e} bands near the magic angle $\theta_{c} \approx 1.1^{\circ}$ also plays an important role in the giant NLH effect studied in this work. This is because the narrow band width of order $\sim 10$ meV near $\theta_{c}$ enables a much stronger $C_3$ breaking by strain than the cases with larger band widths for twist angle $\theta$ away from $\theta_c$.

\begin{figure}
\centering
\includegraphics[width=3.5in]{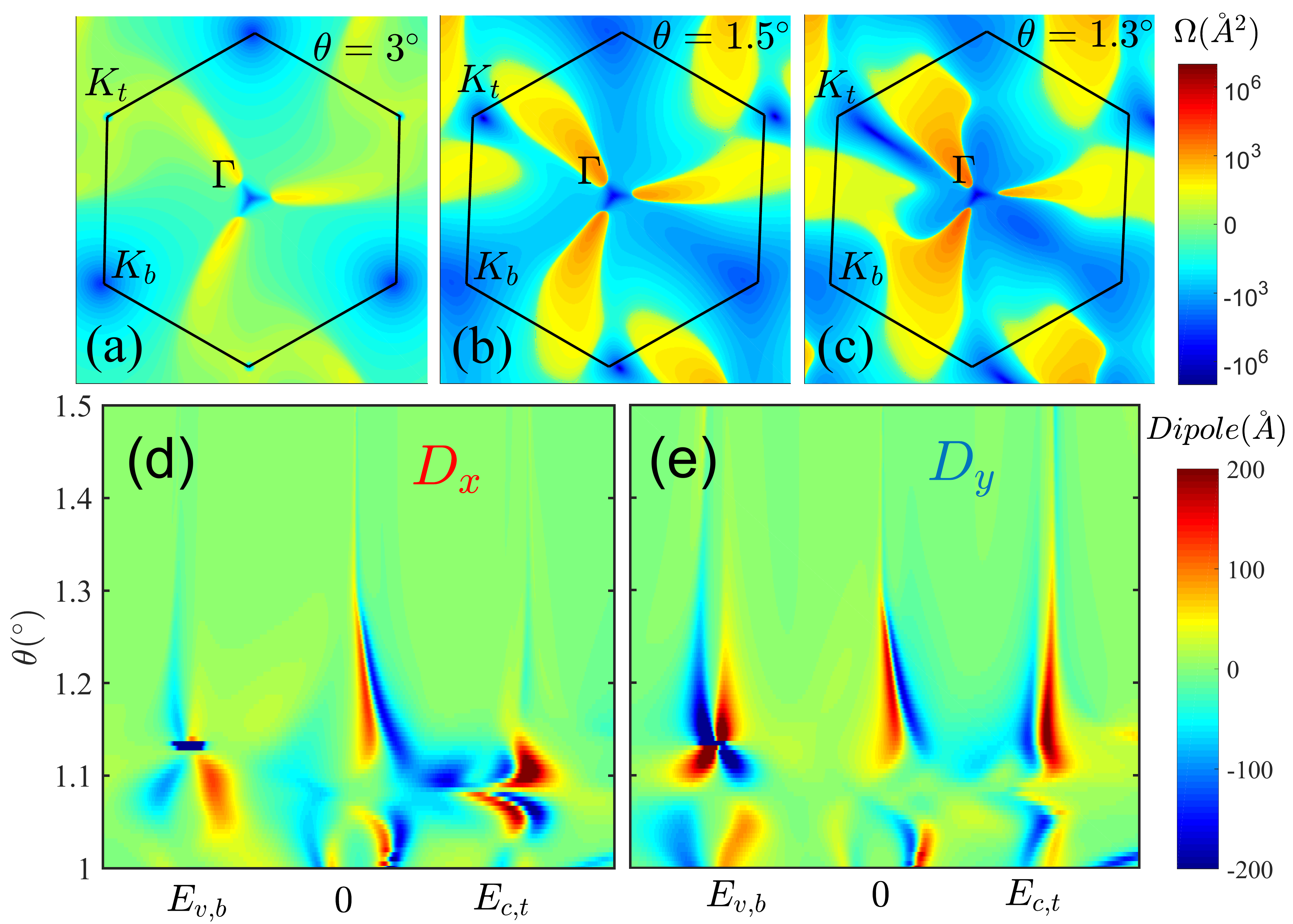}
\caption{(a)-(c) Evolution of Berry curvature in the conduction band as a function of twist angle $\theta$ under fixed strain $\mathcal{E} = 0.1\%$ along the zigzag direction. (d)-(e) Twist angle and gate dependence of $D_x$ and $D_y$ at temperature $T = 2$ K. The Fermi energy $E_F$ is normalized by the average bandwidth $W = (E_{c,t} - E_{v,b}) / 2$ to demonstrate the effect of twist angle on $C_3$ breaking. Here, $E_{c(v),t(b)}$ represents the energy level of the conduction(valence) band maximum(minimum). Giant Berry curvature dipole of order $\sim 200 \; \text{\AA}$ appears near all band edges for twist angle in the range $\theta = 1 \sim 1.3\degree$ near the first magic angle $\theta_{c} \approx 1.1^{\circ}$.}
\label{FIG03}
\end{figure}

To explicitly demonstrate the twist angle dependence of the strain-induced $C_3$ breaking effect, we choose a fixed strain $\mathcal{E} = 0.1\%$ and calculate the Berry curvature profiles at various twist angles $\theta = 3^{\circ}, 1.5^{\circ}, 1.3^{\circ}$ for the conduction band from $+K$ valley as shown in Fig\ref{FIG03}(a-c). At large twist angles $\theta = 3^{\circ}$ (Fig\ref{FIG03}(a)), the distribution of Berry curvature remains highly symmetric under a weak strain. The bandwidth at $\theta = 3^{\circ}$ is typically hundreds of meVs, while the strain-induced gauge potential term $\sim \hbar v_{F} |\bm{A}|$ (see eq.\ref{eq:H_b}) is on the order of several meVs. Thus, the $C_3$ breaking terms induced by strain acts only perturbatively on a TBG at large twist angles. As the twist angle approaches the magic angle $\theta_c$, the moir\'{e} bands become flat as the Fermi velocity near the Dirac points is normalized strongly by the interlayer hopping. The strain-dependent term becomes comparable with the bandwidth and significantly breaks the $C_3$ symmetry, as shown in Fig\ref{FIG03}(b)-(c). This twist angle dependence of the strain-induced band deformation is further illustrated by the complete 2D map of Berry curvature dipole as a function of both twist angle $\theta$ and normalized $E_F$ in Fig\ref{FIG03}(d)-(e). Clearly, the large Berry dipoles of order $\sim 200 \; \text{\AA}$ near band edges appear only within the twist angle range $\theta \approx 1.1^{\circ}$.

\emph{ Conclusion and Discussions.}---In this work, we show that giant NLH effect can arise in a weakly strained TBG, which is characterized by a giant Berry curvature dipole of order $\sim 200 \; \text{\AA}$ exceeding by two orders of magnitude the values of all previously known nonlinear Hall systems \cite{ma2019observation, kang2019nonlinear, you2018berry, zhang2018electrically, du2018band, battilomo2019berry, son2019strain, zhou2020highly, hu2020nonlinear, huang2020giant}. We further pointed out that the giant Berry dipole in strained TBG arises as a general consequence of symmetry breaking induced by $h$-BN substrate and the flatness of moir\'{e} bands near magic angle. Notably, as weak strain of order $1 \%$ is generally found to be present in twisted materials \cite{kerelsky2019maximized,choi2019electronic, xie2019spectroscopic, huang2020giant}, we expect the giant NLH effect in TBG to be easily manifested experimentally.

It is important to note that at special fillings with integer number of electrons/holes per moir\'e unit cell, strong electron-electron interactions drive the TBG into correlated phases \cite{cao2018correlated, yankowitz2019tuning, xie2019spectroscopic}, where the Fermi liquid picture that forms the basis of nonlinear Hall transport theory \cite{sodemann2015quantum} is no longer valid. However, our result of the giant NLH effect in a near-magic-angle TBG still holds for general filling factors. Particularly, we expect the Fermi liquid picture to hold in the regime with giant Berry dipole of order $\sim 200 \; \text{\AA}$ near the band anti-crossing points, where the electron/hole filling factor is very low and far away from the special fillings such as 1/4, 1/2 and 3/4. Thus, complications from electron interactions do not affect our analysis in a qualitative way.


Apart from the NLH effect, we note that the giant Berry curvature dipole found in this work also has important implications for other Berry-dipole-driven phenomena such as nonlinear Nernst \cite{zeng2019nonlinear, zeng2019wiedemann, yu2019topological} and nonlinear thermal Hall effect \cite{zeng2019wiedemann} in TBG. Our results establish TBG as a practical and tunable platform for the study of novel Berry phase effects in twisted materials.

\emph{Acknowledgements.}---The authors thank Wen-Yu He for illuminating discussions. KTL acknowledges the support of the Croucher Foundation and HKRGC through 16324216, 16307117 and 16309718.

\end{document}